\documentclass{aa}  

\usepackage{graphicx}
\usepackage{txfonts}
\usepackage{array,multirow,makecell} 
\usepackage{microtype}
\usepackage[breaklinks, colorlinks, citecolor=blue]{hyperref}
\usepackage{enumitem}
\usepackage{natbib}
\usepackage{hyperref}
\hypersetup{
    colorlinks=true,
    linkcolor=blue,
    filecolor=magenta,      
    urlcolor=cyan,
    citecolor=blue
    }
\usepackage[dvipsnames]{xcolor}

\begin{document}

\title{Implications of SPT and eROSITA cosmologies for \textit{Planck} cluster number counts and t-SZ power spectrum}
\author{G. Aymerich\inst{1,2} \fnmsep\thanks{\email{gaspard.aymerich@universite-paris-saclay.fr}} \and
        M. Douspis\inst{1} \and
        N. Battaglia\inst{1,2,3} \and
        N. Aghanim\inst{1} \and
        L. Salvati\inst{1} \and
        G.W. Pratt\inst{2} \and
        G. Fabbian\inst{1} 
}

\institute{
         Université Paris-Saclay, CNRS, Institut d'Astrophysique Spatiale, 91405, Orsay, France
\and
        Université Paris-Saclay, Université Paris Cité, CEA, CNRS, AIM, 91191, Gif-sur-Yvette, France   
\and
        Department of Astronomy, Cornell University, Ithaca, NY 14853, USA
}

   \date{Received XXX; accepted YYY}

  \abstract{Comparison between cosmological studies is usually performed in a statistical manner at the level of the posteriors of cosmological parameters. In this Letter, we show how this approach poorly reflects the differences between cosmological analyses, when applied to cosmological studies using galaxy cluster abundances. We illustrate this by deriving the implications of the best-fit cosmologies from the recent SPT and eROSITA cluster number counts analyses on the \textit{Planck} thermal Sunyaev-Zeldovich (t-SZ) probes. We first fix the mass calibration, and find that the \textit{Planck} cluster sample would theoretically contain 498 clusters with the SPT cosmology, and 1098 clusters with the eROSITA cosmology, instead of the 439 clusters observed. We then fit the \textit{Planck} number counts to both cosmologies, only varying the hydrostatic mass bias, and find required biases of $0.790 \pm 0.070$ and $0.630 \pm 0.034$ for SPT and eROSITA respectively, instead of the $0.844^{+0.055}_{-0.062}$ derived in \citet{aymerich_cosmological_2025}. Lastly, we compute the expected t-SZ power spectrum obtained from the SPT and eROSITA cosmologies, and compare these to the \textit{Planck} measurement. While the predicted SPT angular power spectrum is in good agreement with the \textit{Planck} measurements, the normalisation of the predicted eROSITA angular power spectrum is two times higher at all scales. These two tests highlight the power of comparing predicted cluster abundances and t-SZ power spectra to measured data in a physically interpretable way.}


\maketitle

\section{Introduction}
The comparison between cosmological studies is often undertaken at the final cosmological parameter level. While this approach has the advantage of allowing comparisons between very different types of analyses, it can render the comparisons not very physically interpretable. This is particularly true for the abundance of large-scale structures in the late-time Universe, which is a source of disagreement between different studies. This problem is known as the $S_8$ tension, where $S_8$ is a combination of the two cosmological parameters given by $S_8\equiv \sigma_8 \sqrt{\Omega_\text{m} / 0.3}$. Historically, most analyses of the late-time Universe \citep[in particular cosmic shear studies such as][]{heymans_kids1000_2021, amon_dark_2022, li_hyper_2023} found a lower $S_8$ value than that predicted by CMB primary anisotropies studies \citep[like e.g.][]{balkenhol_measurement_2023, tristram_cosmological_2024, louis_atacama_2025}. In recent years, the situation has evolved and has now become less clear, with certain late-time studies finding rather high $S_8$ values, compatible with or even higher than the CMB derived values \citep[see e.g.][]{ghirardini_srg_2024, wright_kidslegacy_2025}. In this Letter, we focus on one piece of the $S_8$ tension puzzle, and propose a new approach to galaxy cluster study comparison. Instead of focusing on the best-fit derived $S_8$, we directly estimate the implications of one study's best-fit cosmology on another study's probes, providing a more direct illustration of the differences between analyses. 

We focus on the implications of the best-fit cosmologies from \citet{bocquet_spt_2024a} and \citet{ghirardini_srg_2024} on the thermal Sunyaev-Zeldovich (t-SZ) probes in the \textit{Planck} sky. We choose to compare these studies in particular since the cluster mass calibrations were performed using weak-lensing (WL) shear profiles from Dark Energy Survey (DES) data. This same calibration process was also applied to the \textit{Planck} cluster sample \citep{aymerich_cosmological_2025}, which found that the masses derived for clusters present in both the eROSITA and \textit{Planck} catalogues by the respective best-fit mass calibrations were fully coherent. This means that the differences in final cosmological constraints are not due to the mass calibration \citep[which can shift the final constraints along the $S_8$ direction, see e.g.][]{pratt_galaxy_2019} but are caused by differences in the catalogues or in their modelling. 
We study two observables of the t-SZ sky, the cluster number counts and t-SZ angular power spectrum. For the number counts, we compute how many clusters should have been found in the second cosmological \textit{Planck} catalogue of t-SZ sources \citep[PSZ2,][]{planckcollaborationxxvii_planck_2016} according to the best-fit cosmology of \citet{bocquet_spt_2024a} and \citet{ghirardini_srg_2024} and compare it with the observed PSZ2 catalogue. We also derive the mass bias needed to reconcile the observed \textit{Planck} number counts and the prediction from the best-fit cosmologies of the two studies. For the t-SZ power spectrum, we compare the \textit{Planck} t-SZ spectra derived by \citet{tanimura_constraining_2021} with the expectations from SPT and eROSITA cosmologies using the model of \citet{douspis_retrieving_2022}.

\section{Cluster number counts}
For this study, we use the cluster population model (i.e. the part of the likelihood that takes a set of cosmological parameters and outputs a theoretical prediction of cluster number counts) from \citet{planckcollaborationxxiv_planck_2016}, with the mass calibration from the \textit{Planck}+DES analysis \citep{aymerich_cosmological_2025}, as it was found to be consistent with the mass calibration from the eROSITA analysis. Given the fact that the SPT analysis used the same procedure and DES data, we also expect its mass calibration to be fully coherent with the one we use for our investigation. 

We refer the interested reader to \citet{planckcollaborationxxiv_planck_2016} and \citet{aymerich_cosmological_2024, aymerich_cosmological_2025} for a full description of the population modelling and mass calibration. We note here that the mass calibration used for the latter study relies on a scaling relation obtained from \textit{Chandra}-derived masses corrected by the introduction of a hydrostatic mass bias, calibrated with the DES data. This point is of particular importance, as the value of the mass bias depends on the X-ray telescope used to derive the hydrostatic masses. Indeed, masses derived from XMM-\textit{Newton} data \citep[as used in][]{planckcollaborationxxiv_planck_2016} are on average lower than those derived from \textit{Chandra} data, leading to a $\sim$$15\%$ lower $(1-b)$ value\footnote{The mass bias is parametrized by the $(1-b)$ value, defined as $M_\text{H} = (1-b) M$, where $M_\text{H}$ is the hydrostatic mass and $M$ the true halo mass.} \citep{schellenberger_xmmnewton_2015, aymerich_cosmological_2024}.

\subsection{Fixed mass calibration}
\label{fixed_calib}
The first part of our investigation is to fix the the mass calibration to that of the \textit{Planck}+DES analysis and to predict the expected PSZ2 number counts with the \textit{Planck} population model for the SPT and eROSITA cosmologies. This allows us to highlight the impact of the different $S_8$ values in terms of the number of clusters that should have been detected in the \textit{Planck} data to obtain the same final cosmology as other studies. To do so, we randomly select 200 samples from the posterior of each study and predict the theoretical cluster number counts for every set of cosmological parameters. The mass calibration parameters are randomly sampled from the posterior of the \textit{Planck}+DES analysis. We report the results as median and $1\sigma$ uncertainties.

\begin{figure}[h]
    \centering
    \includegraphics[width=\columnwidth]{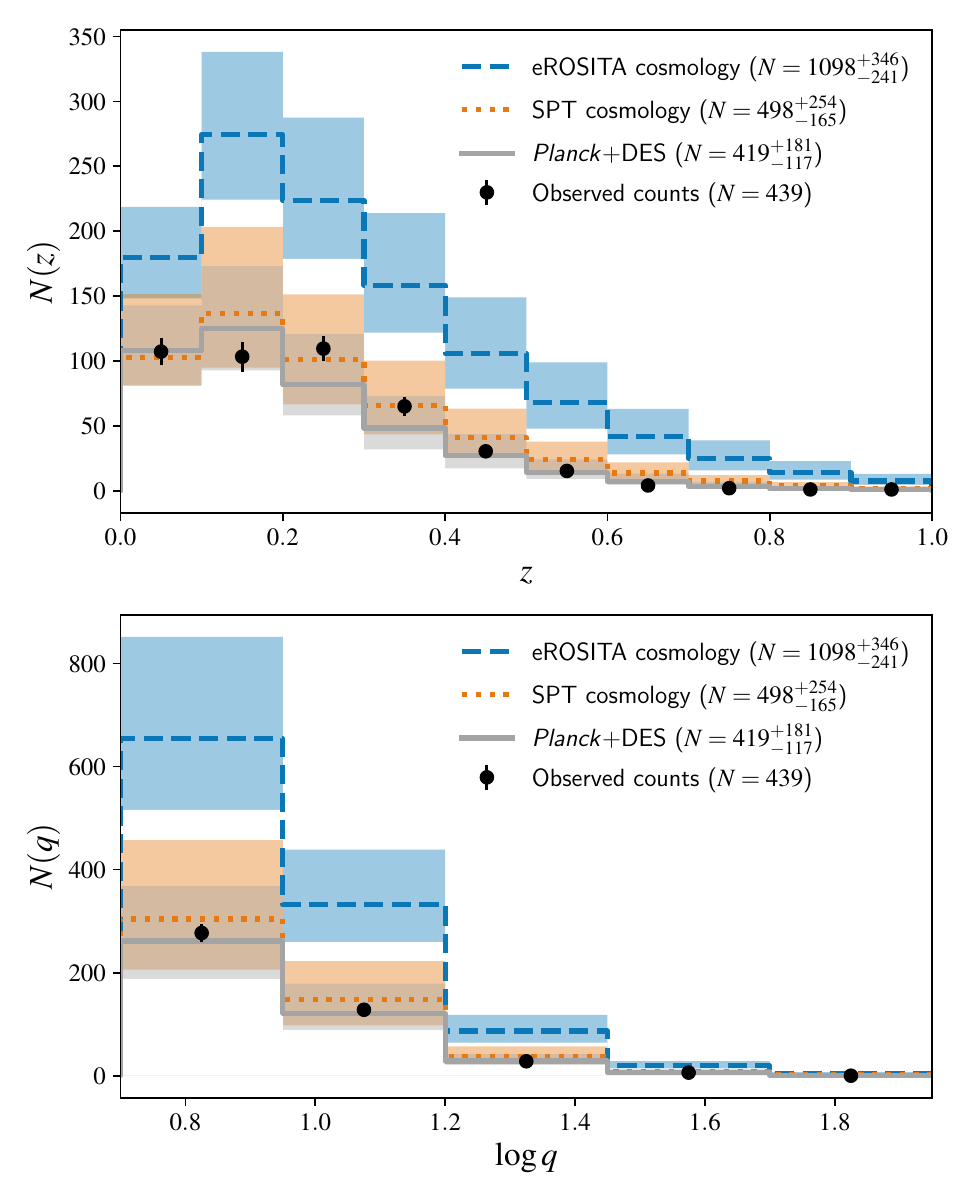}
    \caption{Comparison of the predicted number counts for the \textit{Planck} sample for the cosmologies of eROSITA, SPT, and \textit{Planck}.}
    \label{fig:NC}
\end{figure}

Figure~\ref{fig:NC} shows the observed number counts in the PSZ2 sample and compares them with the expectations for the \textit{Planck}+DES, SPT, and eROSITA cosmologies. We find that the SPT cosmology predicts a slightly higher number of clusters ($\sim$$13\%$) than actually observed by \textit{Planck}, but is fully consistent at the $1\sigma$ level with the \textit{Planck}+DES cosmology. On the other hand, the PSZ2 sample would need to contain 1098 clusters, instead of 439 in the actual sample, in order to obtain the eROSITA best-fit cosmology with the population model from \citet{planckcollaborationxxiv_planck_2016}.

\subsection{Varying the mass bias}
\label{bias_NC}
In Sect.~\ref{fixed_calib}, we focused on the impact of the cosmologies on the number of clusters in the PSZ2 sample while keeping the mass calibration fixed. This was justified by the fact that the mass calibration is probably not the cause of the differences between the analyses, since masses derived by the \textit{Planck}+DES and eROSITA studies are coherent. Here, we change our approach and fit the observed number counts, fixing the cosmology to the SPT and eROSITA posteriors, letting the hydrostatic mass bias vary. To marginalise over the cosmology, we use the likelihood of \citet{planckcollaborationxxiv_planck_2016} and run an MCMC fit of the bias for 200 randomly selected samples from the posterior of each study.

\begin{table*}
  \caption{Mass bias values obtained by fitting the \textit{Planck} number counts with the SPT and eROSITA cosmologies.
  }              
  \label{table:bias}      
  \centering                          
  \begin{tabular}{c c c c}        
  \hline\hline                 
  & \textit{Planck}+DES & SPT & eROSITA \\    
  \hline                        
  $\Omega_\mathrm{m}$ & $0.312^{+0.018}_{-0.024}$ & $0.286 \pm 0.032$ & $0.29^{+0.01}_{-0.02}$\\
  $\sigma_8$ & $0.777\pm 0.024$ & $0.817 \pm 0.026$ & $0.88 \pm 0.02$\\
  $S_8$ & $0.791^{+0.023}_{-0.021}$ & $0.795 \pm 0.029$ & $0.86 \pm 0.01$\\
  $\boldsymbol{(1-b)}$ & $0.844^{+0.055}_{-0.062}$ & $0.790 \pm 0.070$ & $0.630 \pm 0.034$\\ 
  $(1-b)_\mathrm{XMM-like}$ & $0.721^{+0.047}_{-0.053}$ & $0.674 \pm 0.059$ & $0.538 \pm 0.029$\\ 
  \hline                                   
  \end{tabular}
  \tablefoot{The $(1-b)_\mathrm{XMM-like}$ value is obtained by rescaling the \textit{Chandra} bias by the $(1-b)_\mathrm{XMM}/(1-b)_{Chandra}$ ratio from \citet{aymerich_cosmological_2024}. It only  illustrates the bias one would expect to obtain with XMM-\textit{Newton} data, as it might be more familiar in the context of \textit{Planck} clusters.}
\end{table*}

\begin{figure}
    \centering
    \includegraphics[width=\columnwidth]{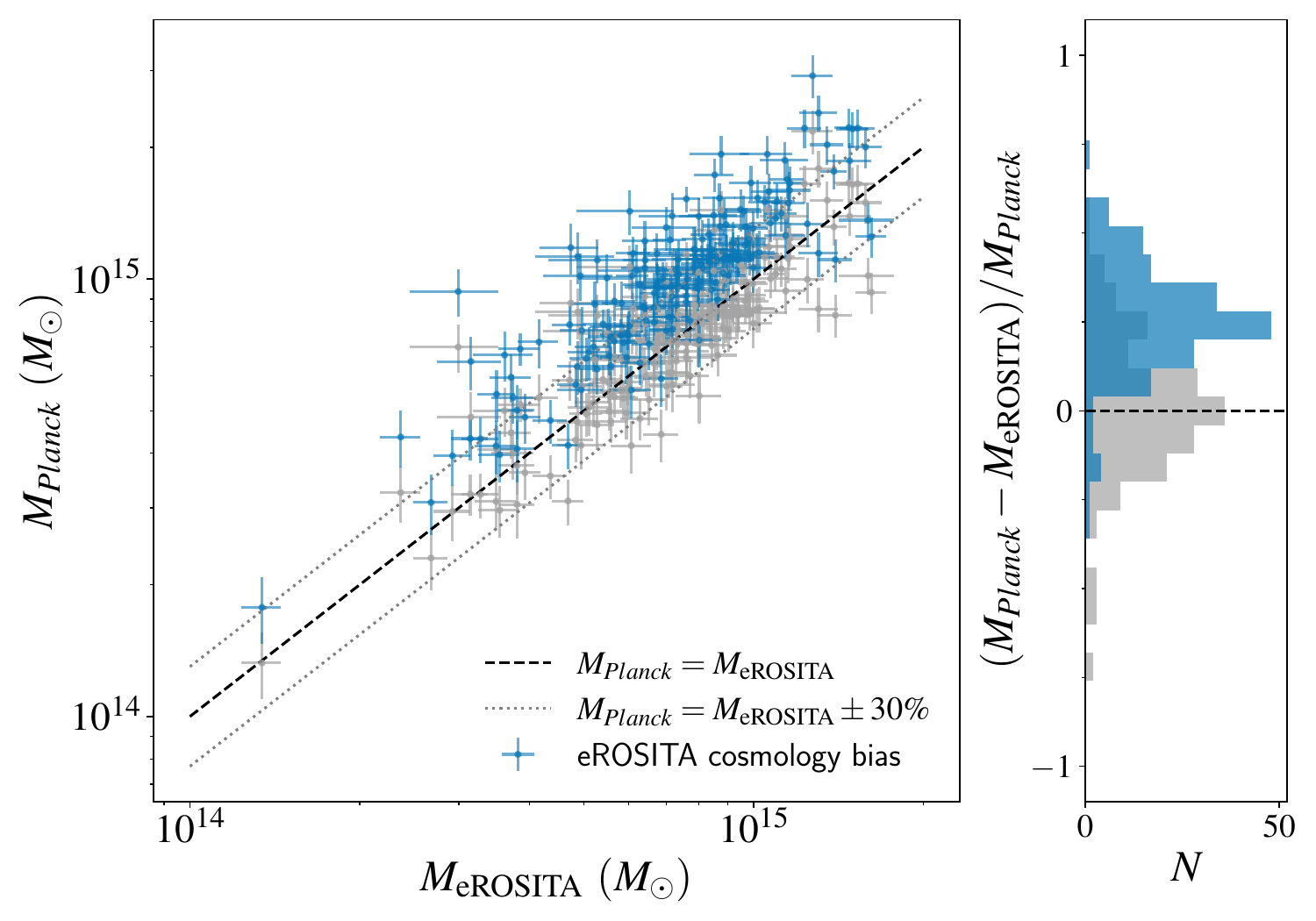}
    \caption{Comparison of eROSITA and \textit{Planck} masses. In grey, \textit{Planck} masses are computed with the best-fit mass bias from the \textit{Planck}+DES analysis. In blue, \textit{Planck} masses are computed with the mass bias required to obtain the eROSITA cosmology with the PSZ2 catalogue and population model.}
    \label{fig:mass}
\end{figure}

Table~\ref{table:bias} presents the hydrostatic mass bias values derived for the SPT and eROSITA cosmologies, and compares them with the hydrostatic mass bias obtained in the \textit{Planck}+DES analysis. Figure~\ref{fig:mass} compares the eROSITA mass \citep[taken from the eRASS1 catalogue,][]{bulbul_srg_2024} and the \textit{Planck} mass for all clusters found in both catalogues. The \textit{Planck} mass is derived either with the mass calibration from the \textit{Planck}+DES analysis (in grey) or with the mass bias obtained for the eROSITA cosmology in this study. We did not include masses obtained for the SPT cosmology bias for clarity, as they are very similar to the \textit{Planck}+DES masses. Obtaining the eROSITA cosmology with the PSZ2 sample would require a very low $(1-b)$ value, especially for a \textit{Chandra} mass bias (the corresponding XMM-\textit{Newton} bias would be $\sim$$0.54$). The masses derived from \textit{Planck} data with this mass bias are not coherent with the masses of the eRASS1 catalogue, unlike the \textit{Planck} masses computed with the DES mass calibration.

\section{t-SZ power spectrum}

We extend our analysis by investigating the effect of the different cosmologies and bias values on the t-SZ angular power spectrum (APS). Using the last \textit{Planck} released maps, \citet{tanimura_constraining_2021} estimated the t-SZ amplitude between $\ell \sim 60$ and $\ell \sim 1000$, marginalising over residual foregrounds (CIB and point sources). The t-SZ APS can be computed for any $\Lambda$CDM cosmological parameter set using the halo model and assumptions on the pressure profile and the hydrostatic mass bias \citep{komatsu_sunyaevzeldovich_2002}. We use the same modelling as in \citet{salvati_constraints_2018} and \citet{douspis_retrieving_2022}, and predict the t-SZ APS for the three cosmological parameter sets derived from the galaxy cluster count analyses discussed above (\textit{Planck}+DES, SPT, eROSITA). We assume a gNFW pressure profile from \citet{arnaud_universal_2010} and the $M-Y_{SZ}$ scaling relation from \cite{planckcollaborationxi_planck_2011}. As these two ingredients are derived using XMM-\textit{Newton} data, we assume, in our modelling of the three angular power spectra, the "XMM-like"  bias found for the \textit{Planck} sample combined with DES data ($(1-b)=0.721$).

\begin{figure}
    \centering
    \includegraphics[width=\columnwidth]{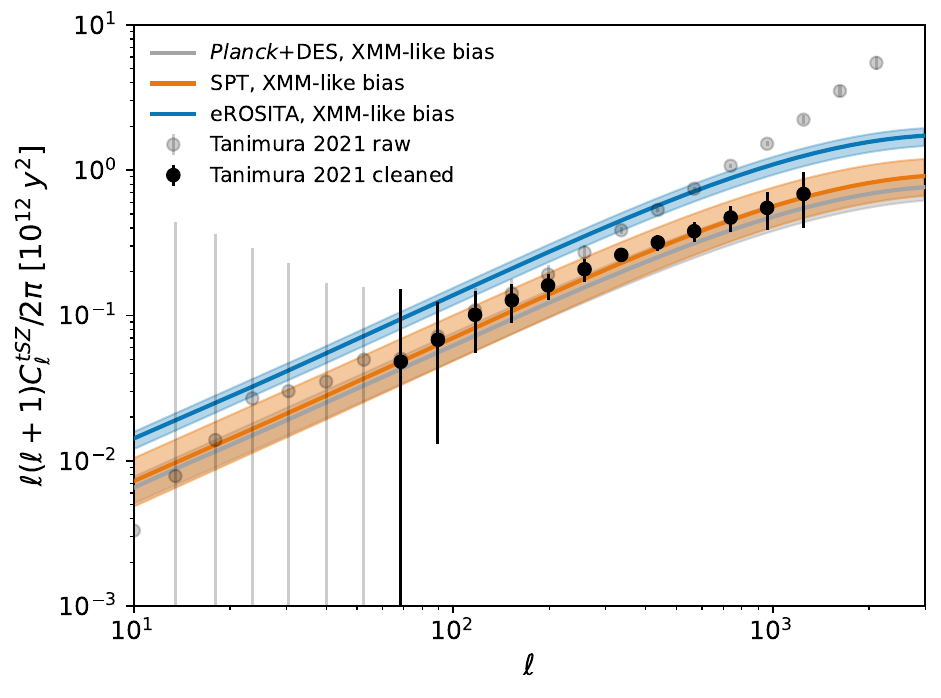}
    \caption{Comparison on t-SZ angular power spectra  assuming the {\it Planck}, SPT, and eROSITA number counts cosmologies. The shaded area takes into account the uncertainties on the cosmological parameters for each analysis (and the trispectrum term of uncertainty for the {\it Planck} power spectrum). The data obtained from {\textit{Planck}} $y$-map  are plotted in black.}
    \label{fig:comp}
\end{figure}

Figure~\ref{fig:comp} shows the three APS with the \textit{Planck} $y$-map APS from \citet[][cleaned in red, raw from the $y$-map in grey]{tanimura_constraining_2021}. 
While the APS obtained using the best model from the {\it Planck} number counts is slightly lower than the data points  of \cite{tanimura_constraining_2021}, 
the SPT APS is in perfect agreement. In contrast, the eROSITA APS is $\sim 2$ times higher in amplitude at all scales, and even exceeds the raw APS. 
The uncertainties associated with the cosmological parameters are plotted in gray, salmon and blue bands for respectively, {\it Planck}, SPT and eROSITA. 

Following Sect.~\ref{bias_NC}, we move to fit the \textit{Planck} t-SZ APS allowing for a free mass bias but fixing the cosmology to eROSITA analyses. The bias needed to reconcile the high value of $\sigma_8$ from eROSITA and the observations is $(1-b)\sim 0.59$. 
We did not perform the analysis for SPT as the APS are already in agreement.

\section{Discussion and conclusions}
In this Letter, we presented the discrepancies between the \textit{Planck}+DES \citep{aymerich_cosmological_2025}, SPT \citep{bocquet_spt_2024a}, and eROSITA \citep{ghirardini_srg_2024} analyses from a new perspective, in terms of t-SZ observables rather than $S_8$ constraints. We first focused on the cluster number counts predicted by the \textit{Planck} population model for the three different cosmologies. We found that the SPT cosmology yields a slightly higher number of clusters, by about $13\%$, but is overall in good agreement with the observed abundance of clusters in the \textit{Planck} sky. In contrast, the number counts predicted with the eROSITA cosmology and the \textit{Planck} population model are about $2.5$ times larger than the observed cluster population. Such a difference is not easily explained, especially since the mass calibrations of the \textit{Planck}+DES and eROSITA analyses are coherent. We also note that both studies use the halo mass function derived in \citet{tinker_halo_2008}. While  uncertainties in the cluster selection functions are expected for all surveys, the \textit{Planck} uncertainties can hardly explain such a large difference \citep[see e.g.][]{gallo_characterising_2024}.

When reversing the problem and trying to fit the observed number counts to the SPT and eROSITA cosmologies by varying the hydrostatic mass bias, we again found that the SPT best-fit cosmology is very compatible with the \textit{Planck}+DES analysis. The required bias of $0.790 \pm 0.070$ is compatible within $1\sigma$ with the value from \citet{aymerich_cosmological_2025}. On the other hand, reconciling the \textit{Planck} number counts and eROSITA cosmology requires a bias value of $0.630\pm0.034$ (or $0.538 \pm0.029$ in the case of the XMM-\textit{Newton}-like bias), in $\sim 3.1\sigma$ tension with the value derived in \citet{aymerich_cosmological_2025}.

\begin{figure}
    \centering
    \includegraphics[width=\columnwidth]{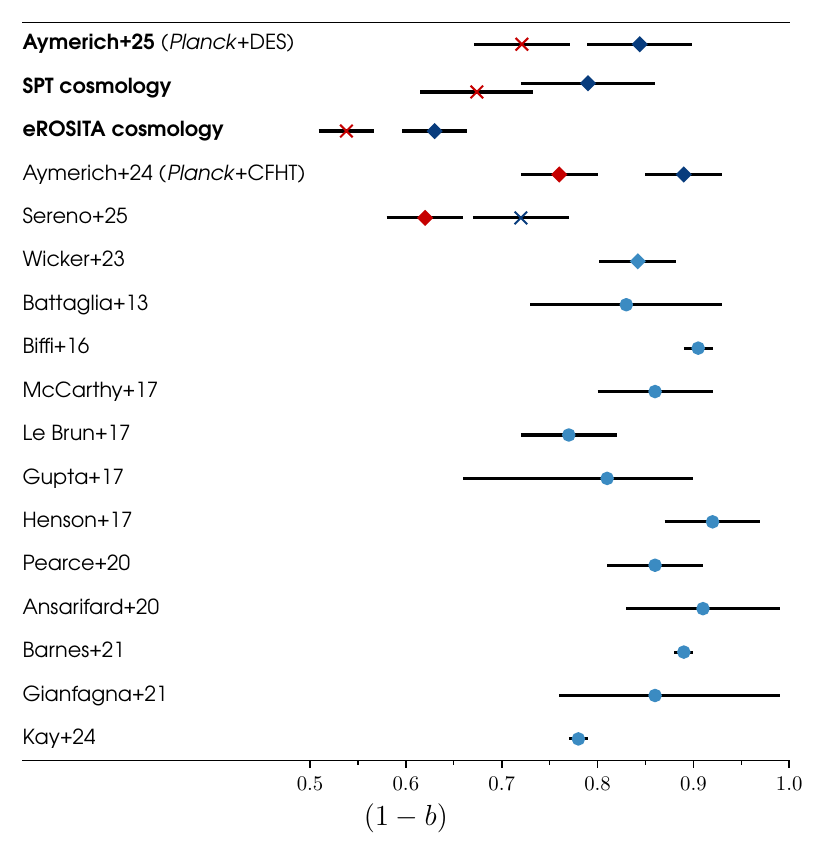}
    \caption{Comparison of the biases obtained in Sect.~\ref{bias_NC} with values from the literature. Two values for a single study indicate that it is subject to the X-ray temperature calibration differences. A dark blue diamond corresponds to a \textit{Chandra} bias and a red diamond to an XMM-\textit{Newton} bias. Crosses indicate \textit{Chandra}-like or XMM-like biases, i.e. biases rescaled to the expected value with data from the other instrument. Light blue points correspond to analyses insensitive to the X-ray temperature calibration problem, with a diamond indicating an observational value and a dot indicating a value derived from simulations.}
    \label{fig:bias_comp}
\end{figure}

Figure~\ref{fig:bias_comp} compares the hydrostatic mass bias values obtained in this study (see Sect.~\ref{bias_NC}) with values from the literature \citep[from][]{battaglia_cluster_2013, biffi_nature_2016, mccarthy_bahamas_2017, lebrun_scatter_2017, gupta_sze_2017, henson_impact_2017, pearce_hydrostatic_2020, ansarifard_three_2020, barnes_characterizing_2021, gianfagna_exploring_2021, wicker_constraining_2023, kay_relativistic_2024, aymerich_cosmological_2024, sereno_chexmate_2025}. The simulation results were taken from the compilation of bias studies by \cite{gianfagna_exploring_2021}, to which the recent FLAMINGO results have been added \citep{kay_relativistic_2024}. The observational results are a compilation of a few recent bias values derived from cluster samples \citep[studies focused on a single object, such as][were not considered]{lebeau_mass_2024, xrismcollaboration_constraining_2025}. This list of results combines many different analyses that differ in their mass definitions and is not meant for direct comparison, but rather to illustrate the range of expected bias values. 

The comparison of the bias values in Fig.~\ref{fig:bias_comp} shows that the bias obtained with the SPT cosmology and the bias derived in \citet{aymerich_cosmological_2025} are within the expected range of hydrostatic bias values. However, the $(1-b)=0.630$ value required by the eROSITA cosmology is very low, and inconsistent with the values derived by most studies, as shown in Fig.~\ref{fig:bias_comp}.

When computing the t-SZ power spectrum for the three different cosmologies, we found that the APS predicted for both SPT and \textit{Planck} cosmologies are in good agreement with the foreground-cleaned APS obtained by \citet{tanimura_constraining_2021}. The APS predicted for the eROSITA cosmology was found to be higher than the foreground-cleaned APS by a factor $\sim$$2$ on all scales. At large scales ($\ell \lesssim 500$), the predicted APS even exceeds the raw APS obtained directly from the \textit{Planck} $y$-map. We found that the bias needed to reconcile the predicted APS with the foreground-cleaned measured APS, while keeping the eROSITA cosmology, is $(1-b)\sim 0.59$. This value is in reasonable agreement with the $(1-b)=0.538 \pm 0.029$ XMM-like bias derived from the number counts fitting in Sect.~\ref{bias_NC}, and lower than what is expected from previous studies of the hydrostatic mass bias (see Fig.~\ref{fig:bias_comp}).

In conclusion, we highlight the tension between recent cluster analyses in terms of observables, namely t-SZ power spectrum and cluster abundance. Instead of comparing $S_8$ values, we showed the discrepancies between the studies in a more physically interpretable way and found that reconciling \textit{Planck}, SPT, and eROSITA cluster cosmologies requires rather extreme assumptions in terms of selection function and/or mass calibration.

\begin{acknowledgements}
GA acknowledges financial support from the AMX program. MD acknowledges the support of the French Agence Nationale de la Recherche (ANR), under grant ANR-22-CE31-0010 (project BATMAN) and the Hubert Curien (PHC-STAR) program. 
NB acknowledges support from the visiting programme of Université Paris Saclay and from CNES for his long-term visit at IAS. The authors acknowledge long-term support from CNES. This research has made use of the computation facility of the Integrated Data and Operation Center (IDOC, \url{https://idoc.ias.u-psud.fr}) at IAS, as well as the SZ-Cluster Database (\url{https://szdb.osups.universite-paris-saclay.fr}).
\end{acknowledgements}

\bibliographystyle{aa}
\bibliography{biblio}

\end{document}